\documentclass[10pt,journal,letterpaper,compsoc]{IEEEtran}
\usepackage{amssymb,amsmath}
\usepackage[dvips]{graphicx}
\usepackage{times}
\usepackage{algorithm}
\usepackage{algorithmic}
\graphicspath{{figures/}}
\usepackage{subfigure}
\usepackage{color}

\definecolor{Blue}{rgb}{0.0,0.0,1.0}
\newcommand{\changed}[1]{#1}
\IEEEoverridecommandlockouts

\begin{document}
\newtheorem{mydef}{Theorem}
\title{Decentralised Learning MACs for Collision-free Access in
WLANs}
\author{Minyu Fang, David Malone, Ken R. Duffy, and Douglas J. Leith
\thanks{The authors are with Hamilton Institute, NUI Maynooth,
Ireland. Work supported by SFI Grants RFP-07-ENEF530, 07/SK/I1216a and HEA's
Network Maths Grant. (email: ken.duffy@nuim.ie).}}
\IEEEcompsoctitleabstractindextext{%
\begin{abstract}
By combining the features of CSMA and TDMA, fully decentralised
WLAN MAC schemes have recently been proposed that converge to
collision-free schedules. In this paper we describe a MAC with
optimal long-run throughput that is almost decentralised. We then
design two \changed{schemes} that are practically realisable, decentralised
approximations of this optimal scheme and operate with different
amounts of sensing information.  We achieve this by (1) introducing
learning algorithms that can substantially speed up convergence to
collision free operation; (2) developing a decentralised schedule
length adaptation scheme that provides long-run fair (uniform) access to
the medium while maintaining collision-free access for arbitrary numbers
of stations.
\end{abstract}
}

 \maketitle

\section{Introduction}
\IEEEPARstart{I}n Wireless Local Area Networks (WLANs), the Medium
Access Control (MAC) protocol regulates access to the communication
channel and plays an important role in determining channel utilisation.
Based on Carrier Sense Multiple Access/Collision
Avoidance (CSMA/CA), the IEEE 802.11 Distributed Coordination
Function (DCF) is the most commonly employed MAC in WLANs. In this
MAC, time on the medium is divided into idle slots of fixed length,
$\sigma \mu s$, and busy slots of variable length during transmissions.
Frames are positively acknowledged to allow retransmission on
failure. In a network with
more than one transmitter, a significant disadvantage of the DCF
is that there is a persistent possibility of collision.
In contrast, Time Division Multiple Access (TDMA) based MACs can
make better use of the radio channel by eliminating collisions.
However, traditional TDMA has drawbacks, typically employing a
central controller that must maintain detailed knowledge of each station
queue occupancy
and their topology, which requires extra exchanges of data.

\begin{figure}
   \centering
   \includegraphics[width=3.1in]{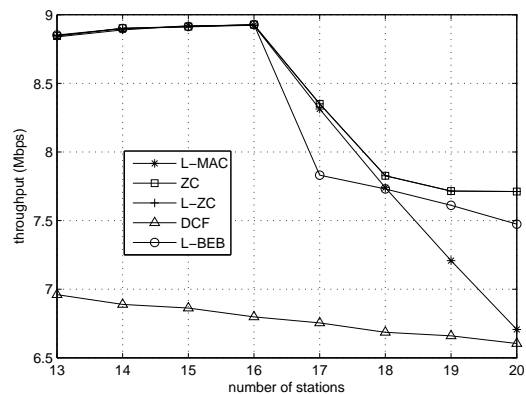}
   \caption{Network throughput vs. number of stations, comparison
   of MACs. Schedule length $C = 16$. Ns-2 simulations. L-ZC overlays ZC.}
  \label{throughput_no_error} %
\end{figure}

New hybrid MAC protocols that retain the best aspects of both TDMA
and CSMA/CA have recently been proposed. For example,
Fig.~\ref{throughput_no_error} shows the throughput performance of
a number of MACs that we will discuss, which can be seen to outperform
DCF by almost 30\% by avoiding collisions.
ZC \cite{Lee08} is a decentralised scheme that
achieves fast convergence to collision-free operation using
information about every MAC slot, not just those where it transmits,
as DCF does. Another collision-free scheme, Learning Binary Exponential
Backoff (L-BEB) \cite{barcelo2008lba} uses a fixed or reselected
random backoff value to achieve collision-avoidance.  Like 802.11's
DCF, it chooses backoff values based on the success or failure of
the last transmission, making it amenable to implementation on
existing platforms. L-BEB converges to collision-free operation
considerably more slowly than ZC.  Other schemes have also been
proposed, see Section~\ref{work} for a brief review.

\begin{figure}
   \centering
   \includegraphics[width=3.1in]{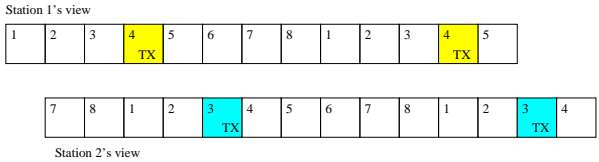}
   \caption{Two stations using a schedule length of $C = 8$, with differing
   views of where the schedule begins and ends, but achieving
   collision-free operation. Note, the slots shown are not fixed
   time PHY slots, but correspond to MAC level slots that will be
   filled by counter decrements, transmissions or collisions.}
  \label{fig:slots}
\end{figure}

Both ZC and L-BEB effectively allow each station to independently
produce a periodic schedule of when to transmit, in terms of MAC
slots, where each slot begins at the point DCF would decrement its
counter, resulting in an idle slot, a successful transmission or a
collision. As the schedules
are periodic and have a length corresponding to
a fixed number of slots, no agreement is required on the labelling
of the slots and the important factor for collision free operation
is that stations transmit periodically but in different parts of
the schedule (see Fig.~\ref{fig:slots}). Consider a CSMA-like
implementation, where stations choose a backoff counter and then
transmit after observing that number of idle slots.  Then periodic
schedules are obtained when each station chooses a fixed backoff
counter equal to the schedule length.

An ideal revision of a TDMA/CSMA hybrid would work with a schedule
with the number of slots equal to the number of active stations and
then instantly converge to a collision-free allocation of stations
to slots. This MAC would ensure there was a successful transmission
in every MAC slot, and so would offer high performance. In practice,
convergence is not instantaneous and the number of active stations
may not be fixed (or even known to all) in a decentralised system.

In this paper, we propose two modifications that can be made
to L-BEB and ZC to provide a good approximation to this ideal hybrid MAC.
\begin{enumerate}
\item
We adapt ideas from a decentralised channel selection algorithm
introduced in \cite{leith2006smd,duffy08}, inspired by learning
automata \cite{Narendra},
to improve convergence times.  In particular, we
propose a fully decentralised Learning MAC
(L-MAC) that uses the same information as L-BEB, but achieves
convergence orders of magnitude more quickly. Similar ideas are also
applied to ZC, and we demonstrate a learning version, L-ZC
provides convergence that is faster than ZC.
\item
In Fig.~\ref{throughput_no_error}, throughput begins to fall when
the number of active stations exceeds 16, the selected schedule length.
When the number of active stations exceeds the schedule length, collisions
are inevitable.  Fortunately, the
quick convergence that is provided by learning allows us to
introduce a mechanism that automatically adapts schedule length.
Using the information available to ZC, we show how the schedule
could be adapted in a centralised way. We then show how this can
be performed in a  decentralised fashion that does not require
agreement between stations while crucially retaining fairness
properties expected of the MAC. This allows MACs to scale to any
number of stations. We call the resulting MACs A-L-MAC and A-L-ZC.
\end{enumerate}

These final algorithms are fully decentralised and do not require information
exchange among transmitting stations or additional control frames
that would increase system complexity. (A-)L-MAC only uses feedback
concerning whether each transmission is successful or not. This
information is already provided by IEEE 802.11 hardware and, thus,
L-MAC can be implemented with relatively minor changes on a
flexible MAC platform. In contrast both ZC and (A-)L-ZC provide
enhanced performance but require additional information on each
slot on the medium, restricting their implementation to future
hardware.

We prove that L-MAC and L-ZC converge to a collision-free
schedule, if one exists.
We determine how to
set the learning parameters of these algorithms. For L-MAC, we use
simulations to choose parameters that offer a balance between
fairness and efficiency.  For L-ZC we provide mathematical analysis
of convergence time that enables analytic optimisation of the
algorithms parameters.

By avoiding collisions, network throughput is significantly higher
than DCF.  In particular, reducing the convergence time to
collision-free operation offers improved performance for delay-sensitive
traffic such as voice, in addition to enhancing throughput
in networks with many station where the 802.11 collision rate is
likely to be large \cite{bianchi2000pai}.  Faster convergence also
allows these schemes to accommodate changing network conditions.
Finally, scalability to networks of any size is enabled by addressing
the fundamental issue of adapting the schedule length in a decentralised
way while still retaining fairness.

The reminder of this paper is organised as follows. Section~\ref{work}
outlines the related work on collision-free channel access methods.
L-MAC and L-ZC are defined in Section~\ref{csma} and appropriate
values for their parameters are identified in Section~\ref{params}.
The schedule length adaptation scheme for optimal long-run throughput
is described in Section~\ref{schedadapt}, along with its practical
decentralised approximations A-L-MAC and A-L-ZC.
Simulation results are provided to illustrate performance
in Section~\ref{result}, where we look
at factors such as performance in the face of imperfect channels
and reconvergence time after network changes.
Section~\ref{conclusion} draws conclusions.  The appendices contain
analytic results regarding the performance of L-MAC and L-ZC.

\section{Related Work}
\label{work}

Z-MAC \cite{rhee2008zmh} is a hybrid protocol that combines TDMA
with CSMA \changed{in wireless multi-hop networks}. Z-MAC assigns
each station a slot, but other stations can borrow the slot, with
contention, if its owner has no data to send; the collision-free
MAC proposed in \cite{busch2004cfm} has less communication complexity.
Both of these MACs experience the same drawback that extra information
exchange beacons are required.  These introduce additional system
complexity, including neighbour discovery, local frame exchange and
global time synchronisation.

A collision-free MAC is introduced in \cite{wang2008cfm} for wireless
mesh backbones. It guarantees priority access for
real-time traffic, but it is restricted to a fixed wireless network
and requires extra control overhead for every transmission. Ordered
CSMA \cite{chen2007occ} uses a centralised controller to allocate
packet transmission slots. It ensures that each station transmits
immediately after the data frame transmission of the previous station.
It has the drawback of requiring a centralised controller with its
associated coordination overhead.

Recently, Barcelo et al. \cite{barcelo2008lba} proposed Learning-BEB,
based on a modification of the conventional 802.11 DCF.  In a
decentralised fashion, it ultimately achieves collision-free TDMA-like
operation for all stations The basic principle of its operation is
that similarly to the 802.11 DCF, stations use a backoff counter and
transmit after observing that number of idle slots.   However, in
Learning-BEB all stations choose a fixed, rather than random, value
for the backoff counter after a successful transmission.
After a colliding transmission, they choose the backoff counter
uniformly at random, as in the DCF. We can
think of this as each station randomly choosing a slot in a schedule,
until they all choose a distinct slot. Arriving at this collision-free
schedule can take a substantial period of time. In particular, when
the number of slots
in a schedule is close to the number of stations, it will take an
extremely long time to converge to collision-free scenario. The authors
of \cite{yong2009incp}
propose a scheme, SRB, that is similar in spirit to L-BEB.

In hashing backoff \cite{starzetz2009hashing} each station chooses
its backoff value by using asymptotically orthogonal hashing
functions. Its aim is to converge to a collision-free state.  One
structural difference from L-BEB \cite{barcelo2008lba} is that
\cite{starzetz2009hashing} introduces an algorithm to dynamically
adapt the schedule length using a technique similar to Idle
Sense \cite{heusse2005iso}. The broad principles of these MAC
protocols are similar
and both have the drawbacks of slower convergence speed to a
collision-free state and lower robustness to new entrants
to the wireless network, relative to our improvements.

\changed{A randomised MAC scheme for wireless mesh networks is
proposed in \cite{yi2010} that also aims to construct a collision-free
schedule. The scheme allocates multiple fixed-length slots in a
fixed-length schedule to satisfy station demands using on-hop
message passing.  If additional sensing information is available,
the authors also show how to improve convergence of the algorithm
through the use of extra state information.}

ZC is proposed in \cite{Lee08}. We can regard ZC as being similar
to L-BEB in that on success it effectively chooses a fixed backoff.
On failure, however, a station looks at the occupancy of slots in
the previous schedule. The station chooses uniformly between the slot
it failed on previously and the slots that were idle in the last
schedule. By avoiding other busy slots, which other stations have
`reserved', ZC finds a collision-free allocation more quickly than
other schemes.

\section{Learning MAC and Learning ZC}
\label{csma}

In this section we consider how learning can be applied to ZC and
L-BEB to improve how quickly they converge to a collision-free
schedule. We describe the scheme for ZC first, as it is more simple.
The scheme for L-BEB is more complex, but offers much greater
improvements in convergence times, without the use of additional
sensing information.

\subsection{The L-ZC protocol}

L-ZC is a modification of the ZC protocol proposed in \cite{Lee08}.
In ZC, each station initially chooses randomly and uniformly from
the all available slots. If it is successful, it chooses
the same slot in the next schedule. Otherwise, it notes the $n_i$
idle slots from the previous schedule and the slot that resulted
in a collision, and for the next schedule it chooses randomly among
these with a uniform probability $1/(n_i+1)$.

In L-ZC we introduce a parameter $\gamma$, that will control the
probability that we choose the same slot after a collision.
\begin{enumerate}
\item Initially L-ZC chooses a slot uniformly in $\{1, 2, \ldots, C\}$.
\item After each schedule, L-ZC updates its choice of slot. If the
	station transmits successfully, or it does not transmit but
	its chosen slot is idle, then it chooses the same slot again.

	If the stations transmission fails, or there is no transmission
	and the station observes a transmission in its chosen slot,
	then the station selects the same slot with probability
	$\gamma$ or chooses one of the $n_{i}$ idle slots with
	probability $(1-\gamma)/n_{i}$.
\item Return to step 2).
\end{enumerate}
The rationale is that different numbers of stations see particular
slots as available for choice, depending on whether a slot was idle,
busy or the chosen slot of a particular station in the previous schedule.
By controlling the weight assigned to collision slots, we are able
to improve convergence times.

L-ZC uses the same information that ZC does.
It needs to know if its own transmission was successful and which
of the previous schedule's slots were idle.

\begin{mydef}
Suppose that all stations employ the decentralised L-ZC. Assuming
that the number of stations $N$ is not more than $C$, for any $\gamma\in(0,1)$
the network converges with probability one in finite time to a
collision-free schedule.
\label{thm:lzc}
\end{mydef}

\begin{proof}
See Appendix.
\end{proof}

\subsection{The L-MAC protocol}
Here we propose a decentralised Learning MAC (L-MAC),
which can be regarded as an evolution of L-BEB \cite{barcelo2008lba}
incorporating ideas from the self-managed decentralised channel
selection algorithm in \cite{leith2006smd}. The primary difference
between L-MAC and L-BEB is that in L-BEB collisions cause memory to
be lost of the current schedule. In contrast, L-MAC keeps some state:
each station that has found a slot that previously
did not have competition is likely to persist with that slot even
after a small number of collisions. To achieve this a probability
distribution is
introduced as internal state for each station. It determines the
likelihood of choosing each slot in a periodic schedule
$\{1,\cdots,C\}$. The advantage of learning is that it introduces
a stickiness that improves the speed of convergence to a collision-free
transmission schedule and facilitates quick re-convergence to a new
schedule when additional stations join an existing network.

L-MAC's slot selection algorithm has a parameter $\beta\in(0,1)$,
the learning strength. For
each station, L-MAC is defined as follows for each station.

\begin{enumerate}
\item
The probability vector $\mathbf{p}(0)$ is initialised at time
$0$ to the uniform distribution,
\begin{align*}
\mathbf{p}(0) =
[p_1(0),\ldots,p_C(0)]
	&=\left[\frac{1}{C},\ldots,\frac{1}{C}\right].
\end{align*}
and a slot $s(0)$ is randomly selected in $\{1,\ldots,C\}$ according to
the probabilities $\mathbf{p}(0)$.
\item
Let $s(n)$ denote the slot selected for transmission in the $n$'th
schedule. We update the probabilities according to success or failure
of a transmission in the slot.

\underline{Success:} If the station has a packet to send and is
successful or
if it has no packet to send and observes the medium to be idle
during slot $s(n)$, then $\mathbf{p}(n+1)$ is set to
\begin{align*}
	p_{s(n)}(n+1) &= 1 \\
	p_{j}(n+1) &= 0
\end{align*}
for all $j\neq s(n)$, $j\in\{1,\ldots,C\}$.
That is, after selecting a non-colliding slot in the schedule, the
station will persist with the same slot $s(n)$ in the following
schedule.

\underline{Failure:} If transmitting in slot $s(n)$ results in a
collision or if the
station has no packet to send and observes the medium to be busy
during slot $s(n)$, then $\mathbf{p}(n+1)$ is set to be
\begin{align*}
	p_{s(n)}(n+1) &= \beta p_{s(n)}(n) \\
	p_{j}(n+1) &= \beta p_{j}(n)+\frac{1-\beta}{C-1}
\end{align*}
for all $j\neq s(n)$, $j\in\{1,\ldots,C\}$. That is, after a failed
transmission, a station reduces the probability that it selects the
same slot again, but it does so in a way that reflects how confident
the station was that the previously selected slot would not result
in a collision.

The station then randomly selects a new slot $s(n+1)$ in the next
schedule using probabilities $\mathbf{p}(n+1)$. In DCF terms, this
amounts to selecting a backoff counter of $C-s(n)+s(n+1)$ slots.
On a success, the backoff counter will always be $C$.

\item Return to step 2).

\end{enumerate}

Before identifying good choices of L-MAC's learning parameter
$\beta$, we state the following theorem that proves that L-MAC
converges to a collision-free schedule if one exists.

\begin{mydef}
Suppose that all stations employ the decentralised L-MAC. Assuming
that the number of stations $N$ is not more than $C$, for any $\beta\in(0,1)$
the network converges in finite time to a collision-free schedule
with probability one.
\label{thm:lmac}
\end{mydef}

\begin{proof}
See Appendix.
\end{proof}

\section{Learning Parameter Choice}
\label{params}

L-ZC and L-MAC both have a learning parameter, $\gamma$ and $\beta$
respectively and a schedule length $C$. In the following subsections
we will identify reasonable values for $\gamma$ and $\beta$. For
L-ZC convergence times are short and our analysis in
Appendix~\ref{proof-l-zc} will show that
convergence times are asymptotically minimised by selecting $\gamma
= 1/(C-N+2)$, where $N$ is the number of stations contending for slots.
For L-MAC we will use simulations to consider factors such as
transient fairness and achievable throughput, as well as convergence
time, ultimately choosing $\beta = 0.95$.

We know from Bianchi's model \cite{bianchi2000pai} that for lower
collision rates the DCF transmission probability will be approximately
$2/(CW_{min}+1)$, close to $1/16$ for the standard value of
$CW_{min}=32$.  Thus, unless otherwise noted, when working with a
fixed schedule length, we set $C = 16$ \changed{so that a converged
station transmits in once in every $16$ slots}. In
Section~\ref{schedadapt}, we will show how the schedule length can
be adapted.

\begin{table}[tb]
\centering
\begin{tabular}[bt]{|c|c|}%
\hline
date rate$=11$Mbps & basic rate=$11$Mbps \\
\hline
PHY header$=24$ bytes & SIFS=$10\mu s$ \\
MAC header$=32$ bytes & DIFS=$50\mu s$ \\
payload=$1020$ bytes & idle slot time$= \sigma =20\mu s$ \\
\hline
\multicolumn{2}{|c|} {header=(MAC header)/(date rate)+(PHY header)/(basic rate)} \\
\hline
\multicolumn{2}{|c|} {ACK=(MAC header)/(date rate)+(14)(8)/(date rate)} \\
\hline
\multicolumn{2}{|c|} {$E_p$=(payload size)(8)/(date rate)} \\
\hline
\multicolumn{2}{|c|} {$T_S$=DIFS+(slot time)+header+$E_p$+SIFS+ACK} \\
\hline
\multicolumn{2}{|c|} {$T_C$=DIFS+(slot time)+header+$E_p$+DIFS} \\
\hline
\end{tabular}
\caption{MAC/PHY values mirroring 802.11b, $E_p$ is the time
spent transmitting payload, $T_S$ is a successful transmission slot
length and $T_C$ is a collision slot length}
\label{parameter}
\end{table}

\subsection{Choosing the collision weight $\gamma$ in L-ZC}

\begin{figure}
 \centering
	\includegraphics[width=3.1in]{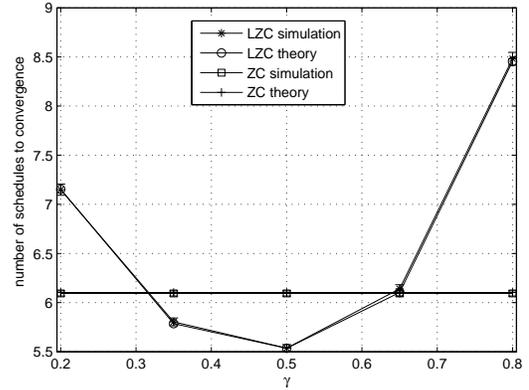}
\caption{Comparison between L-ZC's convergence rate, for a range
of $\gamma$ values, and ZC's convergence rate. $C=16$, $N=16$ stations,
ns-2 simulations and theory}
\label{16_gamma}
\end{figure}

The mathematical analysis of L-ZC in the Appendix allows us to
predict the mean convergence times for different values of $\gamma$
as illustrated in
Fig.~\ref{16_gamma}.  This analysis predicts simulated
times accurately. Based
on our analysis of the subdominant eigenvalue of the Markov chain,
we expect the (asymptotically) optimal value of $\gamma$ to be
$\gamma^* = 1/(C-N+2)$. When $N = C$ the graph confirms that the
shortest convergence time is when $\gamma = 1/2$, and we have found
this asymptotic value seems to match the actual minimum well.

We base our choice of $\gamma$ purely on optimising convergence
time, because it is so short. Reconvergence of ZC/L-ZC to
a collision-free schedule after the addition of new stations amounts
to convergence starting with a smaller number of colliding stations
and free slots.
Thus reconvergence is optimised by optimising convergence. There
will be a period of unfairness during any convergence, but because
of the fast convergence, we believe this should not be a significant
issue.

For a station to choose the optimal $\gamma$, it must know $C-N$,
which corresponds to the number of idle slots when the scheme
converges. This number may be provided by a layer above the
MAC, in which case the exact value can be used. Alternatively, the
station can estimate this value based on the number of idle slots.
For the
remainder of the paper, we assume L-ZC knows the value of $C-N$ and
use $\gamma = 1/(C-N+2)$.

\subsection{Choosing the learning strength $\beta$ in L-MAC}

The learning parameter $\beta$ has an important impact on the
convergence speed, the access fairness while convergence is taking
place, achievable throughput when the network is oversubscribed
(i.e. $N>C$) and reconvergence to collision-free operation after a
change in network conditions. We will see that there is a value for
$\beta$ that ensures convergence is fast while almost optimal
fairness, oversubscribed throughput and reconvergence are achieved.

First, consider the case where there are $N=16$ stations that, in
the terminology of \cite{bianchi2000pai}, are saturated so that
they always have packets to send. The schedule length, $C$, is also
set to $16$. As $N=C$, we are trying to allocate $N$ stations to
exactly $N$ slots, which should be the most challenging case for
the MAC.
Other network parameters are detailed in Table~\ref{parameter}.

\begin{figure}
 \centering
	\includegraphics[width=3.1in]{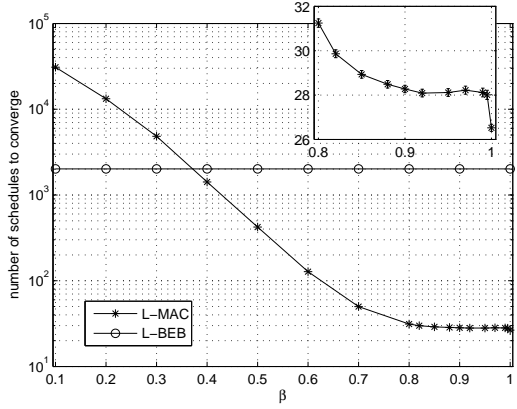}
\caption{L-MAC's convergence time for a range of learning strengths,
$\beta$, and L-BEB on log scale. $C=16$, $N = 16$ stations. The inset
graph shows the detail for $\beta\in(0.8,1)$ on a linear scale.
Ns-2 simulations}
\label{16_beta}
\end{figure}

Fig.~\ref{16_beta} shows the number of schedules required for
convergence versus $\beta$, with $95\%$ confidence
intervals shown based on a Gaussian approximation. Note the larger
graph is on a log scale, while the inset graph is on a linear scale.
It can be seen that larger values of $\beta$ give a smaller number
of schedules (i.e., faster convergence times).  The value of $\beta$
that gives the fastest convergence time is approximately $1.0$.
For $\beta > 0.4$ the time to converge to a collision
free schedule is substantially shorter than that of L-BEB.

A second factor that influences the choice of $\beta$ is its impact
on short-term fairness during convergence to a collision-free
schedule. This is a relevant consideration, as convergence may require tens
of schedules. As we aim for a symmetric sharing of throughput, we employ
Jain's index
\cite{jain1998quantitative,koksal2000analysis,berger2004fairness}
to evaluate fairness.

Fairness is solely a function of the sequence of successful
transmissions.  Consider a network of stations labelled $\{1,\ldots,N\}$.
For each simulation we generate the subsequence of $K$ successful
slots prior to convergence to a collision-free schedule. We record
the sequence of stations that have successful transmissions,
$X_1,\ldots,X_K$, where $X_j\in\{1,\ldots,N\}$.  For each
$m\in\{1,2,\ldots, \lfloor K/N \rfloor\}$, where $\lfloor x \rfloor$
denotes the greatest integer less than $x$, we consider fairness
over windows
of size $w=mN$ successful transmissions. For each station $i$ and window
$k$ of length $w$, we look at the ratio of the actual number of
successes to the number in a perfectly fair allocation:
\begin{align*}
\nu_i(w,k) = \frac{N}{w} \sum_{j=(k-1)w+1}^{kw} 1_{\{X_j=i\}}.
\end{align*}
Then, for each window, Jain's index is given by
\begin{align*}
F(w,k)=\frac{{(\sum_{i=1}^{N}\nu_i(w,k))}^2}
	{N\sum_{i=1}^{N}\nu_i(w,k)^2}.
\end{align*}
Finally we evaluate the empirical average fairness over all windows
in the successful transmission sequence:
\begin{align*}
F(w)=\frac{1}{\lfloor K/w\rfloor}
\sum_{k=0}^{\lfloor K/w\rfloor-1} F(w,k).
\end{align*}
When $F(w)=1/N$ this corresponds to the worst unfairness. Perfect
fairness is obtained when $F(w)=1$. Note that perfect fairness is achieved
by a collision-free schedule and that is why we concentrate on
fairness prior to convergence.

\begin{figure} \centering
   \includegraphics[width=3.1in]{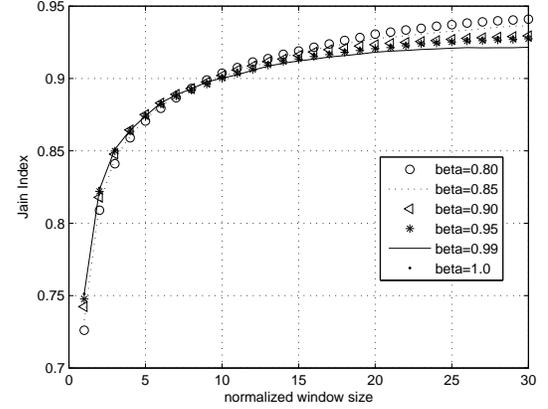}
   \caption{Jain's index vs. normalised window size, $m$, L-MAC
   with different values of $\beta$, $C=16$, 16 stations. \changed{$\beta
   = 1$ values are close to $\beta=0.95$.} Ns-2 simulations}
  \label{jain_full}
\end{figure}

For the data in Fig.~\ref{16_beta}, a
comparison of Jain's fairness index is shown in Fig.~\ref{jain_full}.
In general, we see that smaller values of $\beta$ lead to better
fairness, though the relationship is not monotone, as
0.95 and 1 both offer better fairness than 0.99. We have seen
similar trends in other network configurations, including
oversubscribed networks where $N > C$, (data not shown).

\begin{figure} \centering
   \includegraphics[width=3.1in]{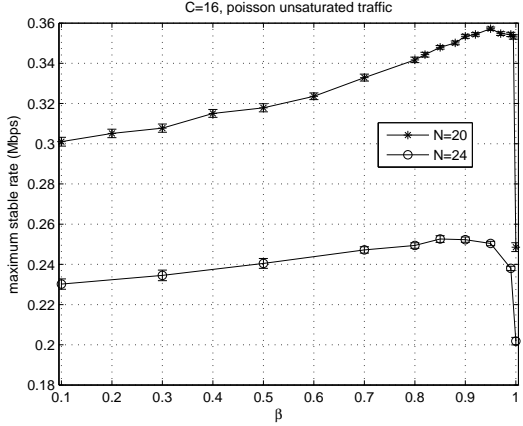}
   \caption{\changed{Achievable stable symmetric rate for different values of $\beta$.
	L-MAC, $C=16$, $N=20$ and $N=24$ stations, ns-2 simulations}}
  \label{rate_region_1}
\end{figure}

Thirdly, we may wish to have reasonable performance when $N > C$ and
there are more stations than slots. We will look at how $\beta$
effects the achievable throughput in this case. It is well-known
that for 802.11-like MACs maximum throughput may not be achieved
when all stations are saturated but may instead correspond to
unsaturated operation \cite{malone2007modeling}. Thus, to find the achievable
throughput, we consider a network with Poisson arrivals at each
station and estimate each station's traffic intensity,
\[
\rho = \frac{\mbox{expected service time}}{\mbox{expected inter-arrival time}}.
\]
Note, both arrival times and service times are stochastic.  To find
the achievable throughput we vary the arrival rate $\lambda$ and
find the largest $\lambda$ that gives $\rho < 1$ for all stations
\cite{asmussen03}. This identifies the stability region when the
network is symmetrically loaded.  For $C=16$ and $N = 20$ and
Fig.~\ref{rate_region_1} shows this upper value of $\lambda$ as
$\beta$ is varied. \changed{We also see that for $N = 24$ stations,
the boundary of the region has similar structure.
This suggests that for an unsaturated network with $N > C$, using
$\beta = 0.95$ gives close to the largest achievable throughput.}

To summarise, convergence time is optimised when $\beta = 1$, but
there is only a small reduction for choosing a value in $(0.9,0.99)$.
In contrast, lower $\beta$ values generally lead to better fairness
before convergence, with values at 0.95 and 1 being comparable.
When we look at the value of $\beta$ that maximises the throughput
region when the network is oversubscribed, we find a value around
0.95 is best, though performance is relatively flat between 0.9 and
1. We have also looked at other metrics, such as reconvergence time
when colliding stations are introduced and we find that there is
an little to separate $\beta$ in a region from 0.75 to 0.95.

Consequently, we suggest that L-MAC use $\beta = 0.95$. This offers
a good compromise between convergence time, fairness and achievable
throughput. We have checked a range of schedule lengths with these
metrics, and find that $\beta = 0.95$ remains an appropriate compromise.

\section{Schedule Length Adaptation}
\label{schedadapt}

As described above, L-ZC and L-MAC use a fixed schedule length $C$.
This can result in reduced performance when $N > C$, as can be seen
in Fig.~\ref{throughput_no_error}. In this section we introduce
an innovative scheme 
allowing schedule length adaptation in a decentralised fashion
while retaining throughput efficiency and fairness. If
information about the number of stations currently contending can
be broadcast to all stations, say by an access point, then stations
can synchronise their schedule length adaptation.

Adapting the schedule length in a decentralised way, while retaining
fairness, is more challenging. If a
decentralised scheme adapts the schedule length independently at
each station, then there is a risk that different stations will use
different schedule lengths (say, because the station is a new entrant
to the network and does not have the same view of the network's
history). This can result in unfairness or even failure to converge
to a collision-free state, because of schedules drifting out of
phase. We will show how to adapt the schedule length independently
at each station, while avoiding problems of unfairness and drifting
phase.

In this section, we begin with an analysis of how the schedule length
impacts on efficiency, where the trade off between idle slots and
collisions is important. We then describe our almost-decentralised
scheme that can provide optimal long-run throughput using the
information available to ZC. We then describe the decentralised
schemes for L-ZC and L-MAC. As the challenges for L-ZC and L-MAC
are similar, we will employ similar schemes for both, however
the L-MAC scheme is more complex because of the more limited information
available to it.

\subsection{The Impact of Schedule Length on Efficiency}
\label{cw_throughput}
As $C$ is the number of available slots for
a collision-free schedule, this is only possible if the
\changed{
number of stations, $N$, is not more than $C$. We will begin by comparing
the long-run throughput when $N$ is less than
}
or greater than $C$.

\changed{
Assume all $N$ stations are are
saturated, and partition $C$ into $C_{suc}$, $C_{col}$ and $C_{idle}$,
}
which denote the number of the successful slots, slots with collisions
\changed{
and idle slots. For 802.11-like protocols,
}
Table~\ref{parameter} shows parameters such as the length of idle
\changed{
and busy slots (e.g. see \cite{bianchi2000pai,malone2007modeling}
to for their derivation). Note that idle slots are an order of
}
magnitude shorter
than successful or collision slots.

When the number of stations $N \leq C$, then,
once we have achieved a collision-free schedule, $C_{col}$ equals
zero and $C_{suc}$ equals $N$. Hence, we get $C_{idle}=C-N$.
\changed{
Then, in the notation of Table~\ref{parameter}, we get a
}
normalised throughput of
\begin{eqnarray}
\label{thr_n_less_cw} S=\frac{NE_p}{N T_S+(C-N)\sigma}.
\end{eqnarray}

\changed{
When $N > C$, we perform an approximate analysis
}
of throughput under an assumption of large $\beta$ for L-MAC or a full L-ZC
schedule with a moderate number of excess stations. We assume 
that each slot will have a single station `stuck'
to it and that the remaining $N-C$ stations are allocated to slots
uniformly randomly with probability $1/C$. The number of slots occupied by
the $N-C$ stations will be the number of slots experiencing collisions,
\changed{
$C_{col}$. This is is a balls-in-bins problem, where we
assign $N-C$ balls to $C$ bins, giving a mean number of
occupied bins of
}
\begin{eqnarray}
\label{e_n_coll}
E(C_{col})= C \left( 1 - \left(1 - \frac{1}{C}\right)^{N-C} \right).
\end{eqnarray}
With this estimate of $C_{col}$ and $C_{suc}=C-C_{col}$, we obtain
\changed{
a normalised throughput of
}
\begin{eqnarray*}
\label{thr_n_more_cw} S=\frac{C_{suc}E_p}{C_{suc}T_S+C_{col}T_C}.
\end{eqnarray*}

\begin{figure}
   \centering
   \includegraphics[width=3.1in]{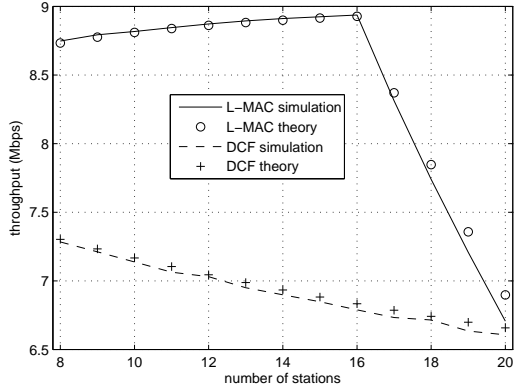}
   \caption{Network throughput vs. number of stations, comparison
   between the theoretical model and the simulation results. $\beta
   = 0.95$. Ns-2 simulations and theory based estimates}
  \label{n_sta_thr_1}
\end{figure}

For example, consider the throughput as $N$ changes and $C$ is fixed
at 16, as shown in Fig.~\ref{n_sta_thr_1}. For comparison, DCF's
throughput is also shown
(the theoretical results for DCF are produced using the well-known model
from \cite{bianchi2000pai}).
\changed{
There is a good match between our predictions and
and the simulation results. Observe that L-MAC's throughput gradually
}
increases as we increase the number of stations $N$ to be the same
as the number of slots. This is because we are eliminating short
idle slots and replacing them with long successful transmissions.
A further increase in $N$ results in a rapid decrease in
throughput. This is because we replace successful slots with
long collision slots. Despite this, L-MAC continues to outperform
DCF until $N=20$ stations. In conclusion, the maximum throughput
\changed{
is achieved when $N = C$, and a slightly smaller throughput
}
is maintained when $N$ is smaller than $C$ as busy slots are of
considerably longer duration than idle slots.

\subsection{Almost-decentralised optimal scheme}

If a station can announce a value of $C$ to be used by the network,
the problem of adapting schedule length is considerably simplified.
Consider a system using L-ZC, where the access point can announce
$C$. The access point can simply observes if the schedule is full,
and if so it can increase $C$ by one. If there are two or more idle
slots $C$ will be decreased by one.

We can easily prove that this adaptation will continue until $C =
N+1$, for if $C < N$ there must be colliding stations, and with
non-zero probability these stations can jump to fill all $C$ slots
in the schedule. Thus we can bound below the probability that $C$
will increase to $N$, when the schedule will be full and then $C$
will increase to $N+1$.  If $C > N+1$ then it is clear that $C$
will decrease, because at least two slots must be free.

This provides $N$
slots filled with transmissions and one idle slot. This idle slot
will allow new entrants to join the network and also from
Section~\ref{cw_throughput}, we know the difference in throughput
between this and $C = N$ will be small. For long-run conditions
with $N$ active stations, this is optimal in the sense that the
maximum number of slots per schedule will be filled with successful
transmissions.

\changed{This scheme is simplistic, but provides us with an example
of how schedule length adaptation can work. In a practical situation,
one might increase or decrease by more than one slot at a time to
accommodate churn in the number of active stations. The threshold
number of idle slots could also be changed, trading a small reduction
in throughput for improved adaptability.}

\subsection{Adaptive schedule length for A-L-ZC}

When choosing a value for $C$ it is better to overestimate
the number of required slots in a schedule. Indeed, Fig.~\ref{n_sta_thr_1}
shows that even with one station too many (i.e. $N=17$), there can
be a greater loss in throughput than having half the slots idle
(i.e. $N=8$).

We will show how to adapt the value of $C$, per station. If stations
operate with
different values $C_i$, two problems may arise. First, stations
are trying to learn a good periodic schedule and so stations'
schedules must not drift with respect to one another. Second, a
station transmits once in every $1/C_i$ slots when a collision-free
schedule is found, so fairness issues can arise.

We address the first problem by using schedules lengths that
all divide evenly into one another. Consequently, when comparing
two stations, the station with the long schedule sees the station
with the short schedule as having claimed a number of fixed slots
within the longer schedule. We use lengths $2^n B$, where $B$ is a
base schedule length. We note that any integer could be used instead
of 2, however using 2 gives the finest granularity.

To address the fairness-related problem, we can
choose to transmit multiple packets in a single slot using a technique
such as 802.11e's TXOP mechanism \cite{ieee80211e03}. Here, a station
transmits multiple
packet/ACK pairs separated by a short interframe space (SIFS). This
time is short enough that other stations observing the medium will
not consider it to have been idle and so backoff processes remain
suspended. Thus we can avoid (long-term) fairness issues by allowing
a station operating at $C_i = 2^n B$ to transmit $2^n$ packets in a
MAC slot.
Short-term fairness issues will be over a time-scale of shorter
than $\max_i C_i/B$ schedules.

This suggests using an MIMD scheme where if a station finds that
the schedule length is too short to accommodate all $N$ stations
it doubles the value of $C_i$ being used. If the schedule length is
much too large then $C_i$ is halved.  It remains to specify a mechanism
that will trigger increases and decreases. As we do
not require the values of $C_i$ to be the
same at all stations to provide fairness, this gives us increased
flexibility in our choices, as
we will not require the MIMD scheme to arrive at a consensus value
of $C$, or even the same mean value.

L-ZC takes advantage of the positions of idle slots in the previous
schedule, and, as in our almost-decentralised scheme, we use this
as trigger for MIMD in A-L-ZC. That is, the adaptive MIMD
scheme that doubles $C_i$ when there are no idle slots remaining
and halves $C_i$ when the number of idle slots is at least half the
schedule. In order to avoid decreasing $C_i$ while L-ZC is converging
and collisions are still ongoing, we wait until we see two consecutive
schedules with the same number of busy slots before we consider a
possible decrease. \changed{The same adaptation scheme can be used with ZC,
and we call the resulting algorithm A-ZC.}

A-L-ZC always achieves collision-free operation with a fixed
number of stations $N$, as A-L-ZC will spread the stations across
idle slots, resulting in the schedule being filled and an increase
in schedule length. This process will stop when there are enough
slots for all stations and each L-ZC instance assigns a collision-free
schedule.

\subsection{Adaptive schedule length $C$ for A-L-MAC}

We being by noting that while L-ZC uses more information than L-MAC,
once converged they behave in a similar manner. Thus, our reasoning for
the $N \leq C$ case above applies directly. While the exact details of
what happens when $N > C$ are different, the broad principles are
similar: as collision slots are longer than idle slots, it will be
more desirable to have idle slots than collision slots.

This suggests that we can again adapt $C$ using an MIMD scheme, but
with different triggers because of the reduced information available
to L-MAC. The trigger we use for doubling $C_i$ is based on $f(C_i)$,
the number of schedules we need for $C_i-1$ stations starting in a
random configuration to have converged with $0.95$ probability,
which can be determined in advance by Monte Carlo simulation. After
arriving at a schedule length of $C_i$, the station checks every
$f(C_i)$ schedules to see if there collisions in that schedule. If
it sees collisions $C_i$ is doubled, otherwise $C_i$ is unchanged.

We expect that reducing $C_i$ will mainly contribute to improving
short-term fairness, unless it is reduced too far, which can result
in significantly reduced throughput. For this reason, we probe with
halving $C_i$ with a frequency that ensures on average we achieve
at least 90\% throughput possible at the current $C_i$ value.
This ensures that if even all transmissions at the
shorter schedule length fail, we will still see the desired 90\% throughput.
In practice, we expect to see even higher throughput.

Note, that because of this probing of shorter schedule lengths,
A-L-MAC will not achieve indefinite collision-free operation unless
$N \leq B$, i.e. the number of active stations can be accommodated
by the base schedule length.  However, we will see in Section~\ref{result}
that the performance of A-L-MAC is close to A-L-ZC, which can achieve
collision-free operation.

\section{Performance Evaluation}
\label{result}

We have implemented these MAC protocols in \emph{ns-2}.
Unless otherwise noted, all
stations are transmitting saturated UDP traffic (with payload
1000 bytes) and a PHY rate of $11$Mbps.  All stations share the
same physical channel, where each station can hear each other and
there are no hidden nodes.
When simulating DCF, parameters are as for 802.11b. All
simulation results are obtained as mean values over repeated
simulations with different seeds. Error bars based on the central
limit theorem are not shown on the graphs as they are on a similar
scale
to the symbols used for plotting points.

We expect results from DCF, L-BEB and L-MAC to be comparable, as
they work with essentially the same information. Likewise, we also expect ZC
and L-ZC to be comparable, because they both leverage extra information
not available to the other MACs. We expect that the adaptive schedule
length schemes (A-L-MAC, A-ZC and A-L-ZC) will show improved
performance when the number of stations is above the base schedule
length. We will see that A-L-MAC offers performance that is comparable
to A-L-ZC in most situations, even though it uses less information.

\subsection{Speed of Convergence}
\begin{figure}
   \centering
   \includegraphics[width=3.1in]{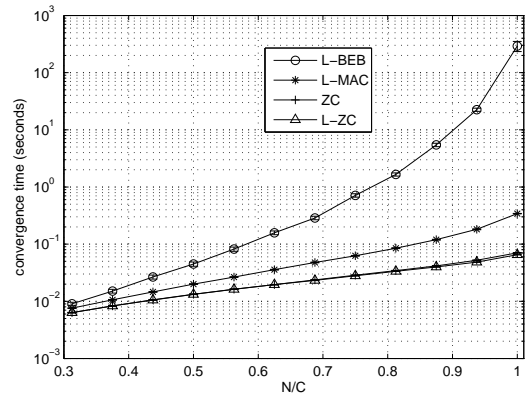}
   \caption{Convergence time vs. $N/C$,
   Comparison between learning MACs, from $N=5$ stations to $N=16$
   stations, $C=16$.
   Ns-2 simulations, error bars too small to be shown}
  \label{beta_nobeta}
\end{figure}

We record the elapsed simulation time\footnote{In previous sections we
presented convergence in terms of the number of schedules
used by the algorithm, rather than real time. These will be related
by the mean slot length during the convergence phase.}
before the
schemes reach a collision-free state (no results are
shown for DCF, as it does not converge). Fig.~\ref{beta_nobeta}
shows this as the ratio $N/C$ is varied. We see that for $N/C < 0.7$
all of the algorithms converge in less than 0.1s.
However as $N/C\rightarrow 1$
we can see the advantages of L-MAC, ZC and L-ZC over L-BEB.
For example, observe that when $N/C = 0.9$ using learning has reduced
the convergence time of 10s for L-BEB to 0.1s for L-MAC. We can
see the advantage of the ZC-based schemes over both L-BEB and L-MAC.
However, it is notable that L-MAC is performing remarkably well for
an algorithm working with less information than ZC and L-ZC.

\subsection{Long-term Throughput}
\begin{figure}
   \centering
   \includegraphics[width=3.1in]{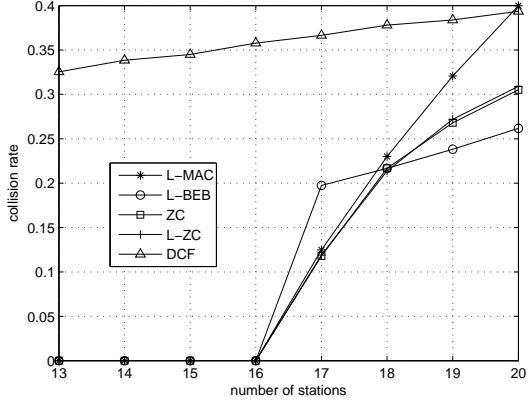}
   \caption{Collision rate vs. number of stations, comparison of MACs,
	ns-2 simulations, $C=16$}
  \label{coll_pro_no_error}
\end{figure}

In Fig.~\ref{coll_pro_no_error} we compare the collision rates of
conventional DCF and the learning schemes with fixed schedule length.
\changed{
These are calculated as the proportion of transmission attempts resulting
in transmissions.
}
L-MAC degrades
gradually with a lower collision rate than DCF's while the number
of stations is between 17 to 19. ZC and L-ZC offer a further reduction
in collision rate. L-BEB's collision probability increases more
quickly when moving from 16 to 17 stations, but then increases more
gradually than L-MAC, ZC and L-ZC, which make
more assumptions about sufficient slots being available.
Fig.~\ref{throughput_no_error} shows the
corresponding results for throughput. This demonstrates that our
learning MACs can achieve good channel utilisation with lower
collision probability than CSMA, even if collisions persist.

\begin{figure}
   \centering
   \includegraphics[width=3.1in]{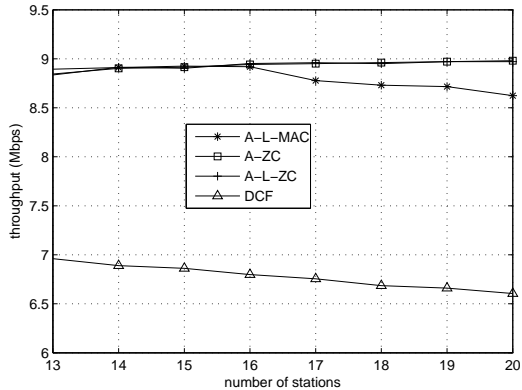}
   \caption{Network throughput vs. number of stations, comparison
   of learning MACs with adaptive schedule length, ns-2 simulations}
  \label{throughput_adaptive_no_error}
\end{figure}

We also investigate the performance of the adaptive schemes for
more than 16 stations. As expected A-ZC and A-L-ZC, achieve a
long-term collision rate of zero. Fig.~\ref{throughput_adaptive_no_error}
shows that A-ZC and A-L-ZC have essentially the same performance,
and A-L-MAC lags only slightly behind. Both adaptive learning schemes
offer substantially higher throughput than that of DCF. Comparing
Fig.~\ref{throughput_no_error} and Fig.~\ref{throughput_adaptive_no_error},
we see how adapting the schedule length allows the schemes to scale
to arbitrary numbers of stations.
While A-L-MAC
shows a slight decline in throughput for $N > 16$, due to probing
shorter schedule lengths, it outperforms
all the non-adaptive schemes (c.f. Fig.~\ref{throughput_no_error}).
A-L-ZC's throughput increases with $N$, as the relative proportion
of idle slots decreases. We have verified this trend out to 50 stations.
We see A-L-MAC still \changed{provides} about 95\% of A-L-ZC's throughput.

\subsection{\changed{Unsaturated Traffic and Delay}}

\changed{
We will assess the behaviour of these protocols in unsaturated
conditions by considering a variable number of stations with Poisson
arrivals at 0.5Mbps. Each station can buffer up to 50 packets.  We
expect that for smaller numbers of stations the network will be
unsaturated and for 20 stations, we expect the network will be
saturated. The network will saturate with different numbers of
stations, because the saturation throughput of the protocols that
we consider varies.  We consider the medium access delay for these
stations, as delay can be an important factor for unsaturated traffic.
}

\begin{figure}
   \centering
   \includegraphics[width=3.1in]{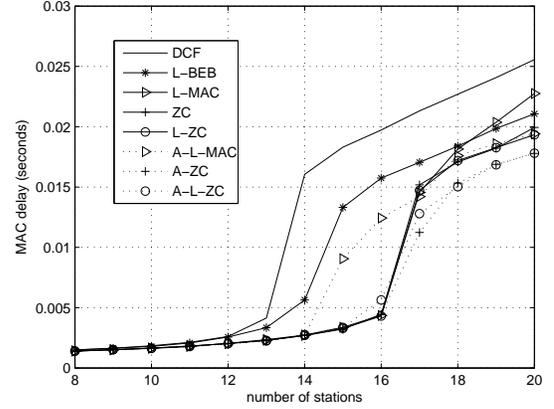}
   \caption{\changed{Mean MAC delay vs. number of stations for unsaturated
   Poisson traffic (0.5Mbps), comparison of MACs with adaptive
   schedule length, ns-2 simulations}}
  \label{mac_delay_all}
\end{figure}

\changed{
Figure~\ref{mac_delay_all} shows the mean medium access delay for
each protocol for $N=8$ to $N=20$ stations. For smaller numbers of
stations, regardless of the protocol, the access delay is similar,
although the learning protocols do have slightly lower delays. As we
increase the number of stations, the access delay increases quickly
as each protocol nears the point where it saturates. This can create
quite large differences in access delay in the region between
saturation for one protocol and another, for example at $N=14$
stations DCF's delay is around 16ms, while the learning MAC's delay
is closer to 3ms. A-L-MAC begins to adapt around $N=15$ stations,
and shows a higher delay than the learning schemes, though still
significantly lower than DCF. Beyond 16 stations we see the advantages
of the adaptive schemes, where access delays are lower than their
non-adaptive counterparts.
}

\subsection{Performance in presence of errors}

In previous graphs we have considered the case of a clean channel
where no packets are lost to noise or interference, and all losses
are due to collisions. A more realistic setting is considered by
introducing errors caused by a fading channel \cite{ni2005sta}. We
consider a simple model where errors are introduced at a particular
rate ($1\%$ and $10\%$). Errors present an interesting challenge
to the learning schemes, because they use transmission failure as
an indication of a slot being occupied.

\begin{figure}
   \centering
   \includegraphics[width=3.1in]{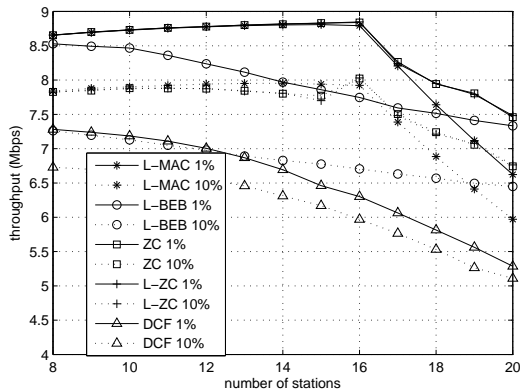}
   \caption{Network throughput vs. number of stations with errors,
   comparison between DCF and MACs with $C=16$h, ns-2 simulations}
  \label{coll_pro_error}
\end{figure}

Fig.~\ref{coll_pro_error} shows the achieved throughputs for the
fixed schedule length learning MACs. We note that DCF's performance
is only slightly degraded by the presence of errors. As all of
L-BEB's state is related to the success of the current transmission,
if suffers quite badly in the presence of errors and its performance
can fall below that of DCF. L-MAC, ZC and L-ZC are more robust to
the presence of errors because their memory is not limited to the
success of a single slot. L-MAC's learning memory will tend to
restore the correct schedule after an error, whereas ZC and L-ZC
can see that other slots have been allocated and do not move to
these slots. A dip in throughput shows that $N=15$ is one of the most
challenging cases for L-ZC and ZC, because there will typically be
one slot available, which several stations will be drawn to in the
case of multiple errors in the same schedule.

We have also investigated the performance of the adaptive schedule
length schemes. As expected, the adaptive schemes offer comparable
throughput to their non-adaptive equivalents for smaller numbers
of stations (data not shown). There is a increase in performance
around 16 stations, similar to that shown in
Fig.~\ref{throughput_adaptive_no_error} where extra slots also help
accommodate churn caused by random losses.

\subsection{Robustness to New Entrants}

\begin{figure}
   \centering
   \includegraphics[width=3.1in]{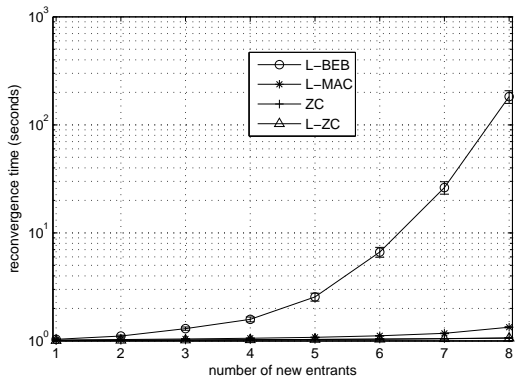}
   \caption{Reconvergence time when $N=8$ stations are in the network
   and a variable number of stations are added, resulting in $N=9$ to
   $N=16$. Ns-2 simulations, $C=16$
   }
  \label{add_time_8_16_error}
\end{figure}

In this section, we briefly consider what happens when the network
has converged, and then more stations are added. We naturally expect
that the improved convergence will extend to quick convergence when
more stations are added to the network. Fig.~\ref{add_time_8_16_error}
shows the time to reconverge to a collision-free schedule after new
stations are added to a collision-free schedule with 8 stations. As
expected, we see rapid convergence, of around one second, even when
8 stations are added to the network at the same time.

\subsection{Coexistence with 802.11 DCF}

This section considers the performance of multiple MAC protocols
used simultaneously on the same wireless channel. All these MACs
are based on the same basic channel-sensing techniques of DCF, so
these MACs should be able to coexist with DCF. Coexistence is a
significant feature of these MACs, because it allows incremental
deployment.

\begin{figure}
   \centering
   \includegraphics[width=3.1in]{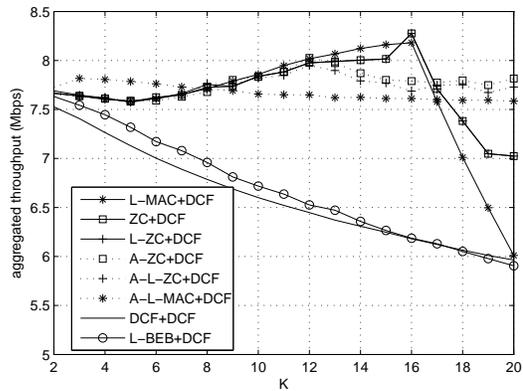}
   \caption{Network throughput for a network $N = 2K$ stations of mixed MACs.
   $K$ of the stations use DCF, and $K$ use another MAC. $C=16$ for MACs
   with fixed schedule length. Ns-2 simulations}
  \label{two_mac_all_thr}
\end{figure}

We consider a scenario where we have $N = 2K$ stations in the network.
Of these stations $K$ use the DCF protocol and $K$ use another
protocol. All the stations are saturated. Fig.~\ref{two_mac_all_thr}
shows the aggregate network throughput achieved as $K$ is varied.
The line for DCF+DCF is our baseline, where all stations use the
DCF protocol. We see that the mixed networks all outperform DCF
alone for $K \leq 16$. For $K > 16$, the throughput of the non-adaptive
learning schemes begins to dip. Up to this point, we expect the
learning schemes to usually allocate one learning station to each
slot, while the DCF stations act as ``noise'', but this is not possible
when there are more than 16 learning stations. We also see
that the adaptive schemes offer slightly lower throughput compared
to the non-adaptive ones just below $K = 16$, because they begin to
increase their schedule length.

\begin{figure}
   \centering
   \includegraphics[width=3.1in]{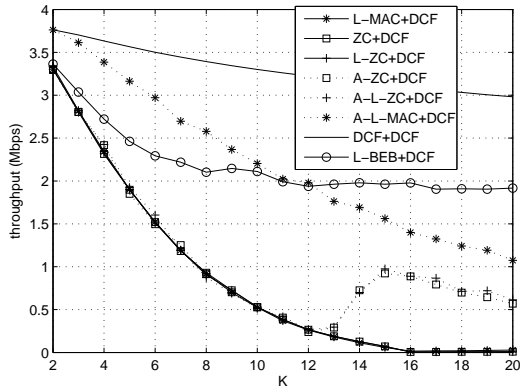}
   \caption{Throughput for DCF stations in network of $N = 2K$ stations
   of mixed MACs.  $K$ of the stations use DCF, and $K$ use another
   MAC. $C=16$ for MACs with fixed schedule length. Ns-2 simulations}
  \label{two_mac_all_thr_dcf}
\end{figure}

The question of how this throughput is shared is also important.
The throughput achieved by the DCF stations is shown in
Fig.~\ref{two_mac_all_thr_dcf}. We see that DCF throughput is
substantially reduced by the presence of stations using
a different MAC, compared to other stations running DCF. Their only
respite is when the adaptive schemes begin to increase schedule
length, making space for the DCF stations to transmit. A-L-MAC
responds to the persistent collisions similarly to a DCF backoff,
and so shares more evenly with DCF.

These results suggest that incremental deployment of these new MAC
protocols would be possible, at the cost of potentially
reduced performance for legacy DCF equipment.

\section{Conclusion}
\label{conclusion}

In this paper we have proposed techniques to improve MACs that
discover collision free schedules. By applying learning, we have
been able to reduce convergence times, improving on L-BEB's convergence
by several orders of magnitude. Using almost decentralised schedule
length adaptation,
we show how L-ZC can lead to an optimal scheme. Crucially, we have
shown how to approximate this in a decentralised way that makes
A-L-ZC and A-L-MAC scalable beyond a fixed number of stations. Of
our two proposed MACs, A-L-MAC uses the same information as DCF,
making it amenable to implementation on existing platforms. A-L-ZC
uses additional information to obtain improved performance, at the
cost of restricting its implementation to more future hardware.
Improvements achieved by L-MAC and L-ZC over DCF and even L-BEB are
substantial, with reduced convergence times, graceful degradation
in the presence of too many stations and improved robustness to
channel errors.

\appendix[Analysis of L-ZC and proof of theorem 1]
\label{proof-l-zc}

\begin{IEEEproof}
The number of colliding stations in next schedule only depends on
current number of colliding stations and the slots they collide on,
hence we build a Markov chain model to study this stochastic
process. We have $N$ stations in the same channel without hidden
nodes, and $C \geqslant N$ per schedule to ensure a collision-free
schedule exists.  We let $N_{(C)}$ be the number of stations
\changed{experiencing a collision in a given schedule}, $n_C$ be
the number of slots with collisions,
and then $n_I=C-N+N_{(C)}-n_C$ is the number of idle slots. We can
immediately establish our result by noting that the probability
that $N_{(C)} > 0$ decreases is lower bounded by
$(1-\gamma)\gamma^{N-1}/C$, the probability that one
station jumps to an idle slot, but all others remain fixed.
\end{IEEEproof}

However, we can give a more refined analysis that enables us to
determine the optimal learning parameter. For each $N_{(C)}$
different configurations of collisions are possible, so we label
these by a sequence $S_{(N_{(C)},i)}=(I_{1},I_{2},\cdots,I_{n_C})$
where $i$ indexes the different states and $I_{j}$ is the number
of stations transmitting in slot $j$. By relabelling the slots, we
only need to consider the case where $I_{j-1}\leqslant I_{j}$ and
we omit slots which have no collision (i.e. $I_j < 2$).  For
example, for two colliding stations, the only possible state is
$S_{(2,1)}=(2)$.  When $N_{(C)}=5$, there are two possible states
$S_{(5,1)}=(5),S_{(5,2)}=(2,3)$.  We denote $\overline S_{N_{(C)}}
:= \{ S_{(N_{(C)},i)} : i \}$ and $\overline S :=
\bigcup_{N_{(C)}=2}^{N}\overline S_{N_{(C)}}$. These sets can be
identified by combinatorial search.

These sequences, $S_{(N_{(C)},i)}$, are the states of our Markov
chain. We add an initial state $IS$ ($N$ stations start to transmit)
and an absorbing state $0$ representing collision-free schedules.
Note that in this discrete-time Markov chain
$S_{(N_{(C)},i)}$ has non-zero probability to transition to state
$S_{(k,j)}$ if $k \leqslant N_{(C)}$ and the state $IS$ has positive
probability to transfer to all states except itself.

Note that the transition probability from $S_{(N_{(C)},i)}$
to $S_{(k,i)}$ is zero if $k > N_{(C)}$, because $N_{(C)}$ is
non-increasing in the next schedule by design.
Assume that $G_{N_{(C)}}$ is a $|\overline S_{N_{(C)}}|\times
|\overline S_{N_{(C)}}|$ matrix of transition probabilities among
states in $\overline S_{N_{(C)}}$ with the same number of colliding
stations.  Considering the state IS and the absorbing state, we
obtain the $(|\overline S|+2)\times (|\overline S|+2)$ full transition
matrix $\Pi$ in upper-triangular block form,

\begin{align}
\Pi= \left[ \begin{array}{ccccccc}
0 & P_{12} & \cdot & \cdot  & \cdot & \cdot & P_{1(2+|\overline {S}|)} \\
0 & G_{N} & \cdot &  \cdot & \cdot & \cdot & \cdot\\
\cdot &   0 & \cdot & \cdot & \cdot & \cdot & \cdot\\
\cdot &  \cdot  & \cdot & G_{N_{(C)}} & \cdot & \cdot & \cdot\\
\cdot &  \cdot & \cdot & \cdot & \cdot & \cdot & \cdot\\
\cdot &  \cdot &  \cdot & \cdot & \cdot & G_{2} & \cdot \\
0 &  \cdot & \cdot  & \cdot &  \cdot & 0 & 1 \\
\end{array} \right].
\label{tr:matrix}
\end{align}

The initial probability measure for all states $\Phi_{(0)} :=
[1,0,\cdots,0]$, at the $n$'th schedule $\Phi_{(n)}=\Pi^{n}\Phi_{(0)}$,
and stationary measure is $[0,\cdots,0,1]$ due to the
absorbing state $0$. The convergence speed depends on the second
largest eigenvalue $\lambda^\ast$ of the transition matrix: the
smaller $\lambda^\ast$, the quicker convergence speed. As $\Pi$ is
a upper triangular matrix, the determinant of $\lambda I-\Pi$ is
the product of determinants of its diagonal entries,
\eqref{eigen_1}.
\begin{align}
|\lambda I-\Pi|=\lambda\prod _{N_{(C)}=2}^{N}|\lambda I-G_{N_{(C)}}| (\lambda-1).
\label{eigen_1}
\end{align}

It is evident that $\lambda_0=0$ and $\lambda_{2+|\overline {S}|}=1$.
In order to get the rest eigenvalues $\lambda$, we will evaluate
the transition matrix $G_{N_C}$, and obtain the largest eigenvalue
of those matrices which is second largest eigenvalue $\lambda^\ast$
of $\Pi$.

Let $\pi_{kl}^{N_{(C)}}$ be the entry of $G_{N_{(C)}}$ corresponding
to the probability of moving from the state $S_{(N_{(C)},k)}=(K_{1},
\cdots)$ to state $S_{(N_{(C)},l)}=(L_{1}, \cdots)$. Let $n_{C}^k$
and $n_{C}^l$ be the number of slots experiencing a collision
in these states respectively.  Consider colliding stations that
choose to remain fixed in the same slot. Since other stations will
have seen that slot as busy, no additional stations will be able
to move into this slot. This if some of the $K_j$ stations remain
fixed, they must correspond to a slot $j'$ with $L_{j'} \leq K_j$.
Let $\Omega \subset \{1, \ldots n_{C}^k\}$ represent slots that
will have some fixed station and let
\begin{align}
M(\Omega) := \left\{ \sigma: \Omega \rightarrow  \{1, \ldots n_{C}^l\}
	: L_{\sigma(j)} \leq K_j,
	\right. \\ \nonumber \left.
	\forall j \in \Omega
	\mbox{ and }
	\sigma \mbox{is one-to-one.} \right\} \nonumber
\end{align}
Note that $M(\Omega)$ may be empty. Let $\{j_1, j_2, \ldots\} :=
\{1, \ldots n_{C}^l\} \backslash \sigma(\Omega)$ be the indices
of collision slots not arising from fixed stations. The number of
stations moving to previously idle slots to produce these collision
slots will be
\[
	m(\Omega, \sigma) := \sum_{j \in \{j_1, j_2, \ldots\}} L_j,
\]
and the number of ways we can choose the idle slots will be
\[
	P(n_I^k, n_C^L - |\Omega|) := \frac{n_I^k!}{(n_I^k-n_C^L+|\Omega|)!}.
\]
So, we may write the transition probability as
\begin{align}
\pi_{kl}^{N_{(C)}}=
	\sum_{\Omega \subset \{1, \ldots n_{C}^k\}}
	\sum_{\sigma \in M(\Omega)}
	\left[ \prod_{j\in \Omega}
	  {K_j \choose L_{\sigma(j)}} \gamma^{L_{\sigma(j)}}\right] \nonumber \\
	\left[ { m(\Omega,\sigma) \choose j_1~j_2~\ldots }
	  \left(\frac{1-\gamma}{n_I^k}\right)^{m(\Omega,\sigma)}\right]
	\frac{P(n_I^k,n_C^l-|\Omega|)}{R},
\label{p_mn_nc}
\end{align}
where $R$ is the number of permutations of the sequence
$S_{(N_{(C)},l)}$ that result in the same state. For particular
$N_{(C)}\in[2,N]$ and $\gamma\in(0,1)$, we can obtain the full set
of states $\overline S_{N_{(C)}}$, obtain the transition matrix
$G_{N_{(C)}}$ based on equation \eqref{p_mn_nc}, and then calculate
the largest eigenvalue $\lambda_{(N_{(C)})}^\ast$ of $G_{N_{(C)}}$.
Then the second largest eigenvalue will be
\begin{align}
\lambda^\ast=\max_{N_C\in[2,N]}[\lambda_{(N_C)}^\ast].
\label{lambda_2}
\end{align}

\begin{figure}
   \centering
   \subfigure[C=N]{\includegraphics[width=1.6in]{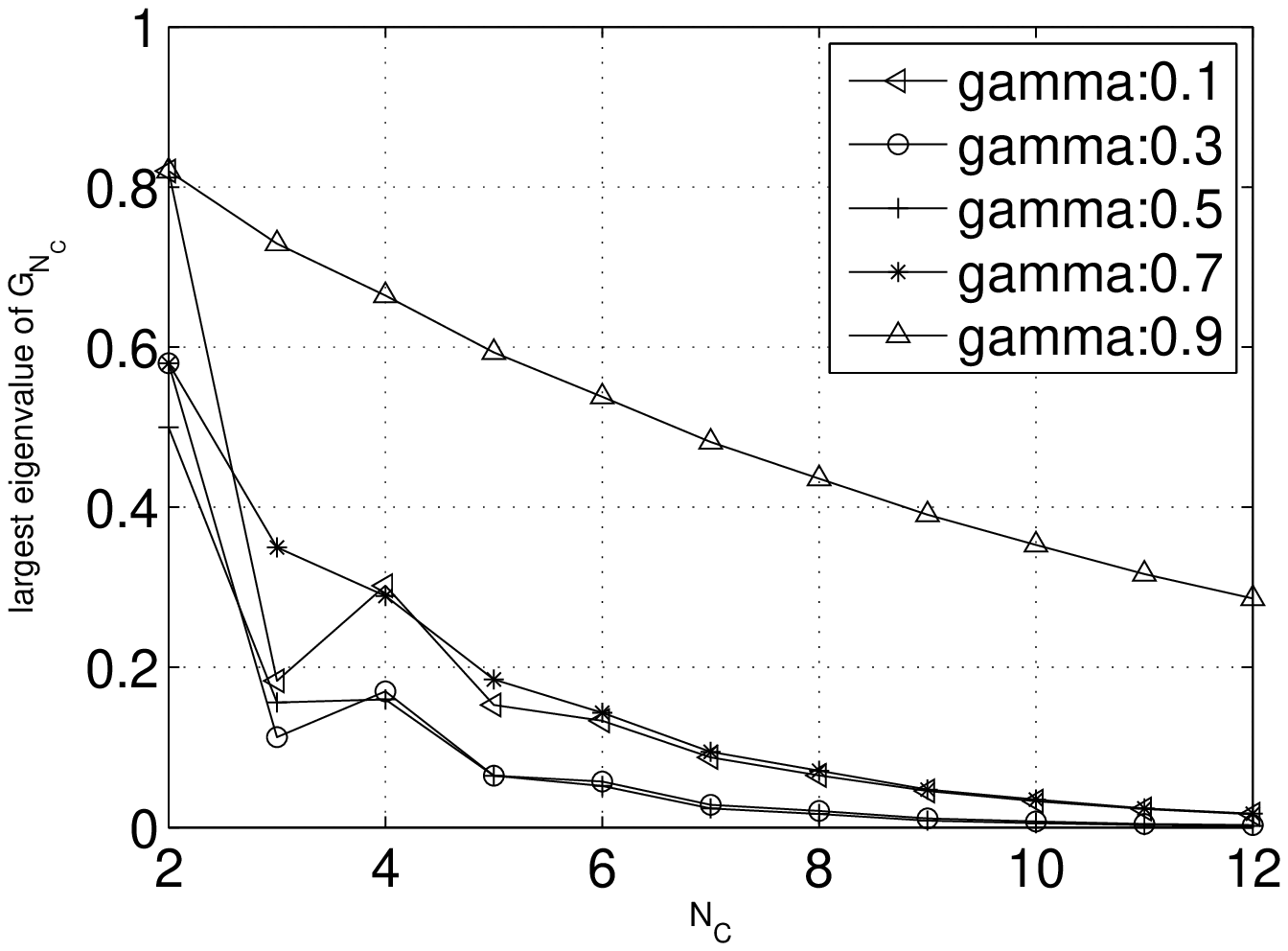}
   \label{eig_0}}
   \hfil
   \subfigure[C=N+2]{\includegraphics[width=1.6in]{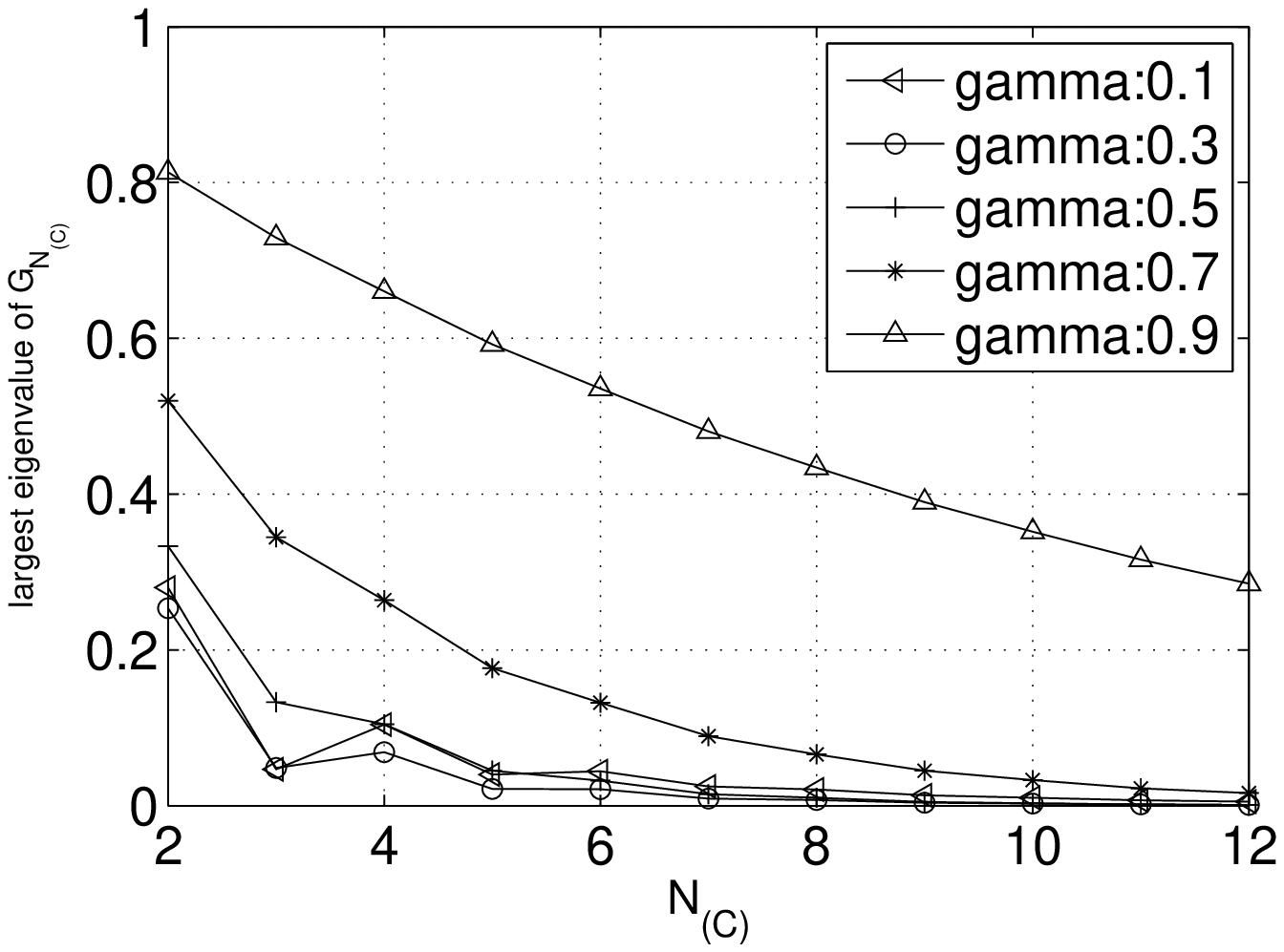}
   \label{eig_2}}
   \caption{Largest eigenvalue vs. $N_{(C)}$ for L-ZC,
   various $\gamma$ values, numerical results}
\end{figure}

Based on this analysis, Fig.~\ref{eig_0} and Fig.~\ref{eig_2} show
the largest eigenvalue of matrix at different $N_{(C)}$ when
$N\leqslant C$. In numerical tests over a range of $\gamma$ values,
we have always observed largest eigenvalue $\lambda^\ast$ is achieved
at $N_{(C)}=2$. Under this assumption we obtain
\begin{align}
\lambda^\ast=\gamma^2+\frac{(1-\gamma)^2}{C-N+1}.
\label{eigen_ast}
\end{align}
Hence, we expect that the minimum $\lambda^\ast$ is obtained by
setting $\gamma=\frac{1}{C-N+2}$.  When $N=C$, $\gamma$ is set at
$0.5$ for the faster convergence speed for L-ZC.

Using this Markov chain, we can also predict the number of schedules
until collision-free
schedule is obtained, assuming that all stations start to transmit
at the same time. 
Let $\Pi_T$ be the transition matrix between all transient states.
We have already obtained the diagonal, $G_{N_{(C)}}$ in equation
\eqref{p_mn_nc} and may obtain most other transitions in a similar
way.  We do have to calculate the first row of $\Pi_T$, representing
transition probabilities from $IS$ into other states $S_{(N_{(C)},i)}$.
If $N-N_{(C)}$ stations choose their own successful $N_{(C)}$ slots,
and $n_C$ slots are chosen from rest $C-N+N_{(C)}$ slots to obtain
the same collision case as $S_{(N_{(C)},i)}$ and the probability
of choosing each slot is initially $\frac{1}{C}$. Thus we get the
transition probability from $IS$ to $S_{(N_{(C)},i)}$ is
\begin{align}
\pi_{IS,S_{(N_{(C)},i)}}= {C \choose {N-N_{(C)}}}P_{(N,N-N_{(C)})} \nonumber\\
{C-N+N_{(C)} \choose {n_C}}/{R}
{N_{(C)}\choose {I_1~I_2~\ldots}}
\left(\frac{1}{C}\right)^{N}
\label{p_IS}
\end{align}
where again, $R$ is number of permutations of $S_{(N_{(C)},i)}$
that result in the same collision state.

Let $\kappa_{(S_{(N_{(C)}),i})}$ denote the number of schedules
elapsed before the network reaching collision-free schedule given
the initial state $S_{(N_{(C)},i)}$, and $\kappa_{(IS)}$ denote the
number of schedules elapsed from state $IS$. Using standard Markov
chain results, the mean number of convergence schedules from initial
state $IS$ is obtained as
\begin{align}
E(\kappa_{(IS)})=[1,0\ldots 0] (I-\Pi_T)^{-1} [ 1,1\ldots 1]^T.
\label{mean_number}
\end{align}
Predictions are shown in Fig.~\ref{16_gamma}.

\appendix[Analysis of L-MAC: proof of theorem 2]
\label{proof-lmac}
\begin{IEEEproof}
By adapting ideas from \cite{duffy08},
we will show that from any state in any two steps of the algorithm,
there is a
probability of convergence that is bounded away from zero. The
probability of selecting a slot can become arbitrarily small if the
station has been colliding on the same slot for many schedules, so
we must construct a sequence of events that avoids this possibility.

Suppose the WLAN consists of $N$ stations. Define
$\mathbf{p}^{(i)}(n)\in[0,1]^C$ to be station $i$'s probability
distribution in the $n$'th schedule and $s^{(i)}(n)\in\{1,2,\cdots,C\}$
to be its slot chosen for transmission.

If we have $s^{(i)}(n)\neq s^{(j)}(n)$, $\forall i\neq j\in\{1,\ldots,N\}$,
then the network has already found a collision-free schedule and
there is nothing to prove.  If, at schedule $n$, there was at least
one collision, then as $C \geq N$, there must be some slot $i^*$, which
has been selected by none of the stations.  At schedule $n+1$, for
any station $k$ colliding at slot $i \neq i^*$ in schedule $n$, the
probabilities of moving to $i^*$ is
\begin{align*}
p^{(k)}_{i^*}(n+1)=\beta p^{(k)}_{i^*}(n)+\frac{1-\beta}{C-1}\geqslant
 \frac{1-\beta}{C-1}.
\end{align*}
Thus the probability that all the stations that collided in schedule
$n$ then, in schedule $n+1$, choose $i^*$ is at least $((1-\beta)/(C-1))^N$.

In schedule $n+2$, the probability a station $k$ that collides in
schedule $n+1$ now picks any slot $j$ is bounded by below by
\begin{align*}
p^{(k)}_j(n+2)=\beta
p^{(k)}_{j}(n+1)+\frac{1-\beta}{C-1}\geqslant \frac{\beta(1-\beta)}{C-1}.
\end{align*}
Since there is at least one non-colliding configuration, the probability
of jumping to this is at least
\begin{align*}
\left(\frac{\beta(1-\beta)}{C-1}\right)^N.
\end{align*}

In summary, no matter what the slot-selection conditions for stations
are in schedule $n$, the probability of schedule $n+2$ being
collision-free, $P(\vec{\mathbf{p}}(n+2)\in A)$,
is bounded below by:
\begin{align*}
	K := 
	\left(\frac{1-\beta}{C-1}\right)^N
	\left(\frac{\beta(1-\beta)}{C-1}\right)^N > 0
\end{align*}
Let $\tau$ be the first time a collision-free schedule is found,
we want to show $P(\tau<\infty)=1$.  At time $2n$, the probability
of arriving at collision-free schedule for the first time is:
\begin{align}
P(\tau\geqslant 2n) \leqslant(1-K)^n. \label{eq:tau}
\end{align}
Thus, as $n\rightarrow \infty$ for any $(1-K)\in(0,1)$, this equation
implies:
\begin{align*}
\lim_{n\rightarrow \infty} P(\tau\geqslant n)
		=\lim_{n\rightarrow \infty} (1-K)^n=0.
\end{align*}
and so $P(\tau<\infty)=1$.
Note that equation \eqref{eq:tau} upper bounds the stopping time
$\tau$ by a geometric distribution and, therefore, all of this
stopping time's moments (mean, variance, etc.) are finite.

\end{IEEEproof}

\bibliographystyle{unsrt}
\bibliography{learning-csma}

\end{document}